\documentclass{article}
\usepackage{amsmath}
\usepackage{amssymb}
\usepackage{graphicx}
\usepackage{amsthm}
\usepackage{algorithm}
\usepackage{algpseudocode}
\usepackage{hyperref} 
\usepackage{times}
\usepackage[a4paper, margin=1in]{geometry} 
\usepackage{xcolor}
\usepackage{comment}
\usepackage{booktabs}

\usepackage{pgfplots}
\pgfplotsset{compat=1.18}
\usepackage{xcolor}
\definecolor{cDM}{HTML}{1F77B4}          
\definecolor{cGRAM}{HTML}{FF7F0E}        
\definecolor{cLIKE}{HTML}{17BECF}        
\definecolor{cORACLE}{HTML}{2CA02C}      
\definecolor{cEST}{HTML}{D62728}         
\definecolor{cLMMSE}{HTML}{9467BD}       

\usepackage{tikz}
\usepackage[utf8]{inputenc}
\usepackage{amsmath, amssymb, bbm}
\usepackage[]{enumitem}
\usepackage{wrapfig}
\usepackage{extarrows}
\usepackage{subcaption}
\usepackage[dvipsnames]{xcolor}
\usepackage{hyperref}

\newcommand{\bC}{\mathbb{C}}

\newcommand{\mH}{\mathbf{H}}
\newcommand{\mI}{\mathbf{I}}

\newcommand{\mQ}{\mathbf{Q}}
\newcommand{\mR}{\mathbf{R}}

\newcommand{\mW}{\mathbf{W}}
\newcommand{\mX}{\mathbf{X}}
\newcommand{\mY}{\mathbf{Y}}
\newcommand{\mZ}{\mathbf{Z}}

\newcommand{\tmH}{\tilde{\mathbf{H}}}
\newcommand{\tmY}{\tilde{\mathbf{Y}}}
\newcommand{\tmR}{\tilde{\mathbf{R}}}
\newcommand{\tmU}{\tilde{\mathbf{U}}}
\newcommand{\tmV}{\tilde{\mathbf{V}}}
\newcommand{\tmsigma}{\tilde{\mathbf{\Sigma}}}

\title{GRAM-DIFF: Gram Matrix Guided Diffusion for MIMO Channel Estimation}
\author{Xinyuan Wang and Krishna Narayanan\\
Department of Electrical and Computer Engineering,\\
Texas A\&M University,\\
College Station, TX 77843, U.S.A\\
Email: merewxy@tamu.edu, krn@tamu.edu}
\begin{document}

\maketitle

\begin{abstract}
We propose GRAM-DIFF, a Gram-matrix-guided diffusion framework for semi-blind multiple input multiple output (MIMO) channel estimation. 
Recent diffusion-based estimators leverage learned generative priors to improve pilot-based channel estimation
\cite{fesl,arvinte2023mimo_channel_est_using_diff,zilberstein2024joint,ma2024diffusion_based_channel_est,zhou}; but they do not exploit second-order structural information estimated from data symbols. 
In practical systems, the channel Gram matrix can be estimated from received symbols and it provides realization-level information about channel subspace structure.
The proposed method integrates a pretrained angular-domain diffusion prior with two complementary guidance mechanisms:  a novel Gram-matrix guidance term that enforces second-order consistency during the reverse diffusion process, and
likelihood guidance from pilot observations. 
Signal-to-noise ratio (SNR)-matched initialization and adaptive guidance scaling ensure stability and low inference latency.
Simulations on 3GPP and QuaDRiGa channel models demonstrate consistent normalized mean-squared error (NMSE) improvements over deterministic diffusion baselines, achieving 4 to 6 dB SNR gains at an NMSE of 0.1 over the baseline in \cite{fesl}. The framework exhibits graceful degradation under coherence-time constraints, smoothly reverting to likelihood-guided diffusion when data-based Gram estimates become unreliable.
\end{abstract}

\section{Introduction}
\label{sec:introduction}
Multiple-input multiple-output (MIMO) channel estimation is a central enabling technology for 5G and emerging 6G wireless systems, directly impacting beamforming, spatial multiplexing, link adaptation, and interference coordination in massive antenna deployments. As base stations scale to tens or hundreds of antennas, accurate channel estimation becomes both more critical and more challenging. Two concurrent trends are particularly consequential. First, in massive MIMO regimes, the available transmit power is distributed across an increasingly large number of spatial dimensions, leading to low per-antenna signal-to-noise ratios. In such regimes, effective estimators must leverage the intrinsic structure of the channel and exploit prior information in a principled manner, rather than relying solely on pilot observations. Second, massive MIMO systems impose stringent computational constraints and approaches that vectorize the matrix-valued channel and attempt to learn or estimate high-dimensional statistical models quickly become infeasible. These considerations motivate channel estimation frameworks that preserve the native matrix structure of the MIMO channel, scale gracefully with array size, and integrate data-driven learning in a way that respects known physical and statistical sufficiency principles.

Classical pilot-based approaches focus on directly estimating the channel matrix $\mH$ from known training symbols. These methods implicitly treat the channel as the primary object of inference and rely heavily on pilot overhead to resolve all channel degrees of freedom.
Most classical approaches to MIMO channel estimation rely on explicit statistical models for the channel matrix $\mH$. In pilot-based estimation, classical estimators such as the linear minimum mean-squared error (LMMSE) estimators are optimal only under restrictive assumptions, such as Gaussian channel distributions with known second-order statistics. 
In practice, however, the true channel statistics may be shaped by array geometry, propagation environment, mobility, and hardware impairments, resulting in distributions that are neither Gaussian nor easily parameterized.

Recent advances in learning-based channel estimation have demonstrated that deep neural networks can outperform classical estimators by implicitly learning complex channel statistics from data. Specifically, diffusion models have been trained on channel realizations and then $\mH$ has been estimated only from pilots by treating the channel estimation problem as an inverse problem \cite{arvinte2023mimo_channel_est_using_diff,ma2024diffusion_based_channel_est,fesl,zhou,zilberstein2024joint}. 

In typical communication systems, pilots are not transmitted in isolation; 
they are transmitted in conjunction with data symbols.
Even though data symbols are unknown at the receiver, the received signal corresponding to the data reveals some information about $\mH$ which can be used to perform semi-blind channel estimation.
In particular, the Gram matrix $\mH \mH^{\sf H}$ can be estimated from the received signal.
In signal processing, subspace-based and semi-blind estimation methods that exploit second-order statistics of the received signal to estimate the range or correlation structure of the channel have been studied extensively \cite{Tong1998Tutorial,medles2001semiblind}. 
In \cite{jagannatham2006whitening}, a semi-blind algorithm based on decomposing $\mH$ as $\mH = \mW \mQ^{\sf H}$,
estimating $\mW$ from the received signal and using pilots only to estimate $\mQ^{\sf H}$ is proposed and it is shown to be highly effective in massive MIMO settings.
Indeed, information-theoretic results show that in the absence of full channel state information, schemes that estimate $\mH$ based only on may be suboptimal \cite{hassibi2003much,ngo2025noncoherentmimocommunicationstheoretical}.

However, there have been very few works that combine generative modeling with semi-blind channel estimation. 
Joint data detection and channel estimation with diffusion models is considered in \cite{zilberstein2024joint}; however,
Gram matrix information is not used in the estimation process in \cite{zilberstein2024joint}. A very recent paper \cite{weisser2025semi} considers the use of variational autoencoders for semi-blind channel estimation, but does not consider diffusion models. 

In this paper, we bridge these two lines of research by introducing a Gram-conditioned, diffusion-based channel estimation framework for single-user MIMO systems. 
As in \cite{fesl}, the diffusion model is trained offline on representative channel realizations to capture the true, possibly highly-structured, distribution of $\mH$. 
At inference time, channel estimation is posed as a conditional generation problem, where the diffusion model is conditioned on {\em both} (i) an estimate of the instantaneous Gram matrix $\mH \mH^{\sf {H}}$ obtained from the data portion of the received signal, and (ii) pilot observations that provide additional, complementary information about the channel.

Unlike classical Bayesian estimators that rely on analytically convenient priors, or end-to-end neural estimators that ignore known physical structure, the proposed method uses diffusion models to learn rich, data-driven priors while explicitly conditioning on physically meaningful information about the channel extracted from the received signal. To the best of our knowledge, the use of a learned diffusion prior conditioned jointly on pilot observations and an estimated instantaneous Gram matrix has not been previously explored for MIMO channel estimation.

\section{System Model and Problem Formulation}
We consider a multiple-input multiple-output (MIMO) single carrier system with $N_T$ transmit antennas and $N_R$ receive antennas.
At each time instant, a vector of $N_T$-QAM symbols are transmitted and a vector of $N_R$ complex values are received. 
We will refer to these as vector symbols to clearly distinguish them from the QAM or complex received values.
In the considered setup, data is transmitted in frames which contain $N_p$ pilot vector symbols and $N_d$ information or data vector symbols.
Thus, a frame $\mX$ is a $N_T \times (N_p + N_d)$ matrix of transmitted QAM symbols. 
Let $\mX_p \in \mathbb{C}^{N_T \times N_p}$ and $\mX_d \in \mathbb{C}^{N_T \times N_d}$ denote submatrices of $\mX$ that correspond to the pilot and the data parts and hence, $\mX = [\mX_p \mX_d]$.
Let $\mH \in \bC^{N_R \times N_T}$ denote the complex channel matrix, which is the object of primary interest in this paper.
The received matrix $\mY$ is related to the transmitted matrix of symbols according to 
\begin{equation}
\label{eqn:sysmodel1}
    \mY = \mH \mX + \mZ,
\end{equation}
where $\mZ$ is a matrix of noise samples which are i.i.d complex Gaussian random variables with zero mean and variance $\sigma^2$ per component. 
In this paper, we restrict our attention to the case when $\mH$ remains constant over the transmission of the entire frame.
The received matrix can be split into two matrices corresponding to the pilots and data vector symbols as follows
\begin{align}
    \label{eqn:Hp1}
    \mY_p & = \mH \mX_p + \mZ_p, \\
    \label{eqn:Hd1}
    \mY_d & = \mH \mX_d + \mZ_d.
\end{align}
We consider the case when $\mH$ is unknown at the transmitter and the vector of transmitted symbols at any time $t$ is a vector of i.i.d QAM symbols \cite{weisser2025semi,zilberstein2024joint}. 
We also consider the case of orthogonal pilots where $\mX_p$ is an orthonormal matrix and hence, $N_p = N_T$.
The considered system model accurately models a single carrier system with flat fading and no time-selectivity;
however, this model is also applicable to sub bands of an orthogonal frequency division multiplexing (OFDM) system with frequency selective fading and no time selectivity and multi-user settings.
Extension to other cases will be considered later.

The main problem we address in this paper is the channel estimation problem of estimating $\mH$ given the received matrix $\mY$ and the pilot matrix $\mX_p$. 
In its baseline form, the proposed estimator operates using the pilot observations $\mathbf{Y}_p$ alone and leverages learned structural priors on the channel to produce accurate estimates even at low per-antenna SNR. Importantly, when additional information becomes available - specifically, when the channel Gram matrix $\mathbf{H}\mathbf{H}^{\sf H}$ can be estimated from the data part $\mathbf{Y}_d$, the same framework can seamlessly incorporate this side information to further refine the channel estimate without the need for training. 
Thus, the proposed method gracefully improves with increasing side information, yielding consistent gains when Gram estimates are reliable, while maintaining strong performance in their absence.

The presented approach and results are most interesting in the case when the prior distribution of $\mH$ is complicated.
We use a pretrained diffusion model that is trained to generate samples of $\mH$ and add guidance terms to the diffusion process to improve the channel estimate. 
We first briefly discussion some preliminaries and prior work before stating our novel contribution.

\section{Preliminaries and Notation}
\label{sec:preliminaries}
For any $N_R \times N_T$ matrix $\mH$, let $\tilde{\mH}$ denote the two-dimensional Discrete Fourier Transform (DFT) of $\mH$ given by 
\begin{equation}
    \label{eqn:2D-DFT}
    \tilde{\mH} = \mathbf{\Phi}_{N_R} \mH \mathbf{\Phi}_{N_T}^{\sf T},
\end{equation}
where $\mathbf{\Phi}_N$ is the 1-D DFT matrix of size $N \times N$ representing a $N$-point DFT. 
We assume that the rows and columns of $\mathbf{\Phi}_N$ are normalized. 
Since the columns (and rows) form a set of orthonormal vectors,
$\mathbf{\Phi}_N^{\sf H} \mathbf{\Phi}_N = \mathbf{\Phi}_N \mathbf{\Phi}_N^{\sf H} = \mathbf{I}_{N}$.

The channel matrix $\mH$ in \eqref{eqn:sysmodel1} is naturally specified in the spatial domain since $\mH_{i,j}$ is the channel coefficient between the $i$th receive antenna and the $j$th transmit antenna.
When the transmit and receive antennas are arranged in a uniform linear array, $\tilde{\mH}$ represents the channel in the angular domain. 
It is well known that typical wireless channels have (sparse) representations in the angular domain that are favorable to learning algorithms. 
This property was successfully leveraged in \cite{fesl} to design diffusion models for channel estimation.
We follow their approach and work with the angular representation in this paper as well.

Let $\mR := \mH \mH^{\sf H}$ and $\tilde{\mR} := \tilde{\mH} \tilde{\mH}^{\sf H}$ denote the Gram matrices of $\mH$ and $\tilde{\mH}$, respectively. The two Gram matrices are related through a similarity transformation given by
\begin{align}
    \label{eqn:Grammatrices}
    \tilde{\mR} = \tilde{\mH} \tilde{\mH}^{\sf H} & = 
    \mathbf{\Phi}_{N_R} \mH \mathbf{\Phi}_{N_T}^{\sf T}
    (\mathbf{\Phi}_{N_T}^{\sf T})^{\sf H} \mH^{\sf H} \mathbf{\Phi}_{N_R}^{\sf H} \\
    & = \mathbf{\Phi}_{N_R} \mR \mathbf{\Phi}_{N_R}^{\sf H}.
\end{align}

Throughout the rest of the paper, capital boldface letters represent matrices in the spatial domain and the corresponding variables with tilde refer to matrices in the angular domain. 

\section{Related Prior Work}

Score-based and diffusion-based generative models have been introduced for MIMO channel estimation as a powerful data-driven prior in recent years.
In \cite{arvinte2023mimo_channel_est_using_diff} channel estimation is performed via posterior sampling using score-based generative models, where the posterior gradient is constructed by combining an analytically derived likelihood term with a learned prior score, and samples are drawn using Langevin dynamics. While flexible, this approach requires iterative stochastic sampling and a large number of updates.
In \cite{ma2024diffusion_based_channel_est} the likelihood information is incorporated into the reverse diffusion process. Their method improves estimation accuracy but still relies on reverse sampling with a large number of diffusion steps, leading to high computational complexity.
To reduce inference complexity, \cite{fesl} proposed a deterministic diffusion-based channel estimator by interpreting the noisy observation as an intermediate diffusion step matched by SNR. By truncating the reverse diffusion process and avoiding stochastic resampling, their method achieves low-latency inference and is shown to asymptotically converge to the conditional mean estimator.
Another important contribution of \cite{fesl} is to demonstrate that by performing the reverse diffusion in the angular domain, a low-complexity convolutional neural network suffices to denoise the channel observations.
More recently,~\cite{zhou} extended diffusion-based channel estimation to general high-dimensional linear inverse problems, moving beyond denoising-based formulations by explicitly incorporating likelihood information into the diffusion process. Although highly general and applicable to various channel models, the approach incurs increased computational cost due to per-step likelihood gradient evaluation and matrix operations.

None of the aforementioned papers exploit the fact that the Gram matrix $\mH \mH^{\sf H}$ can be estimated accurately from $\mY_d$ in some cases.
In \cite{jagannatham2006whitening}, a whitening based approach is proposed where the channel $\mH$ is written as $\mH = \mW \mQ^{\sf H}$ and $\mW$ is estimated from $\mY_d$ and then a maximum-likelihood estimate of $\mQ$ is formed from $\mY_p$.
They show that significant gains in the mean-squared error can be obtained by this approach especially when $N_R > N_T$.
However, they do not consider a Bayesian setting where prior information about the distribution of $\mQ$ is obtainable from a dataset.

Very recently, in \cite{weisser2025semi}, semi-blind MIMO channel estimation using variational auto-encoders that leverage generative priors has been proposed; however, this paper does not use a diffusion model. 
In \cite{zilberstein2024joint}, semi-blind MIMO channel estimation using diffusion models is considered and a diffusion model is learned to sample from the joint distribution of the channel and the data given the observations; 
however, the Gram matrix is not used as guidance in \cite{zilberstein2024joint}.

\section{Proposed Algorithm}

Our proposed algorithm is a posterior sampling algorithm using a pretrained diffusion model trained to sample $\tmH$ in the angular domain in conjunction with appropriate guidance from $\mY_p$ and an estimate of $\tilde{\mR}$. 
The angular domain sample $\tmH$ finally is mapped back to the spatial domain via an inverse Fourier transform.
We explain the steps in the algorithm in detail below.

\subsection{Training}
We begin with a training dataset $\mathcal{H} = \{\mH_1, \mH_2, \ldots, \mH_M\}$ that contains $M$ channel matrices in the spatial domain. 
The dataset is transformed to a dataset in the angular domain $\tilde{\mathcal{H}} = \{\tmH_1, \ldots, \tmH_M\}$ by taking the 2-D DFT of each of the data points (matrices).
As shown in \cite{fesl}, using a generative model for $\tmH$ permits the use of simple convolutional neural networks in each step of the diffusion model (denoising step) for sampling from the posterior. 
We use a pretrained diffusion model from \cite{fesl}. 
In the next two sections, we describe how the received signal is processed to obtain guidance. 

\subsection{Pilot-based Preprocessing}
\label{sec:pilot}
Consider the part of the received signal corresponding to the pilots given by
\begin{equation}
\mY_p = \mH \mX_p  + \mZ_p,
\qquad
\mZ_p \sim \mathcal{CN}(\mathbf{0}, \sigma^2 \mathbf{I}).
\end{equation}

We right-multiply the received pilot observation by \(\mathbf{X}_p^{\sf H}\),
which yields
\begin{equation}
\mathbf{Y}_p \mathbf{X}_p^{\sf H}
=
\mathbf{H}\mathbf{X}_p \mathbf{X}_p^{\sf H}
+
\mathbf{Z}_p \mathbf{X}_p^{\sf H}
= \mH + \mZ_p \mX_p^{\sf H}.
\end{equation}
Note that we only consider orthogonal pilots and hence, $\mX_p \mX_p^{\sf H} = \mI$.
Let $\tilde{\mY}_p$ and $\tilde{\mZ}$ denote the 2-D DFT of $\mathbf{Y}_p \mathbf{X}_p^{\sf H}$ and 
$\mathbf{Z}_p \mathbf{X}_p^{\sf H}$, respectively i.e., $\tilde{\mY} \triangleq \Phi_{N_R}  \mY_p \mX_p^{\sf H} \Phi_{N_T}^{\sf T}$
and $\tilde{\mZ}_p \triangleq \Phi_{N_R} \mathbf{Z}_p \mathbf{X}_p^{\sf H} \Phi_{N_T}^{\sf T}$, then we have the simplified form
\begin{equation}
\label{eqn:additivemodel1}
\tilde{\mY}_p = \tmH + \tilde{\mathbf{Z}}_p, 
\qquad 
\tilde{\mathbf{Z}}_p \sim \mathcal{CN}(\mathbf{0}, \sigma^2 \mathbf{I}).
\end{equation}
Note that $\tilde{\mathbf{Z}}_p$ is obtained from 
$\mZ_p$ through a sequence of orthonormal transformations 
and hence, $\tilde{\mathbf{Z}}_p \sim \mathcal{CN}(\mathbf{0}, \sigma^2 \mathbf{I})$.
These operations remove the dependence on the pilots and convert the original linear observation model in the spatial domain into an additive-noise form in the angular domain, effectively recasting the channel estimation task as a denoising problem amenable to diffusion-based generative modeling.
The signal-to-noise ratio (SNR) associated with \eqref{eqn:additivemodel1} is given by $\text{SNR}(\tilde{\mathbf{Y}}_p) = \frac{1}{\sigma^2}$.

We assume that $\tmH$ is preprocessed such that each component of $\tmH$ is normalized to have unit variance. In this case, each component of $\tilde{\mathbf{Y}}_p$ has a variance given by $(1+\sigma^2)$. 
Dividing both sides of \eqref{eqn:additivemodel1} by $(1+\sigma^2)$, we get
\begin{equation}
    \label{eqn:additivemodel2}
    \frac{1}{\sqrt{1+\sigma^2}}\tilde{\mY}_p = \frac{1}{\sqrt{1+\sigma^2}} \tmH + {\frac{\sigma}{\sqrt{1+\sigma^2}}}\tilde{\mathbf{W}}_p, 
\end{equation}
where $\tilde{\mathbf{W}}_p = \frac{1}{\sigma}\tilde{\mathbf{Z}}_p \sim \mathcal{CN}(\mathbf{0}, \mathbf{I})$ is a Gaussian random matrix with standard normal entries.
Note that in the normalized model, the quantity on the LHS has unit variance.

\subsection{SNR-adaptive Reverse Process}
\label{sec:snr}

We consider a variance-preserving diffusion process with a predefined noise schedule
$\{\beta_t\}_{t=1}^T$, and define
\begin{equation}
\alpha_t = 1 - \beta_t, \qquad
\bar{\alpha}_t = \prod_{s=1}^t \alpha_s .
\end{equation}
Under this formulation, the forward process is given by 
\begin{eqnarray}
\label{eqn:forwardprocess}
\tmH_t & = & \sqrt{1-\beta_t}\,\tmH_{t-1}
+
\sqrt{\beta_t}\,\boldsymbol{\varepsilon}_t,
\qquad
\boldsymbol{\varepsilon}_t\sim\mathcal{N}(\mathbf{0},\mathbf{I}),\\
\tilde{\mathbf H}_t & = &
\sqrt{\bar{\alpha}_t}\,\tilde{\mathbf H}_0
+
\sqrt{1-\bar{\alpha}_t}\,\boldsymbol{\eta_t},
\qquad
\boldsymbol{\eta_t}\sim\mathcal N(\mathbf 0,\mathbf I),
\end{eqnarray}
which holds for any $t\in\{1,\dots,T\}$. Here, $\boldsymbol{\varepsilon}_t$ denotes the independent Gaussian noise injected
at each diffusion step.
The forward process also admits an equivalent closed-form reparameterization, where each diffusion state $\tmH_t$ can be expressed using an auxiliary Gaussian variable $\boldsymbol{\eta_t}$. This representation is mathematically equivalent to the step-wise formulation with independent Gaussian noises $\{\boldsymbol{\varepsilon}_t\}_{t=1}^T$.

Here, each component of $\tmH_t$ is normalized to have unit variance
where each diffusion step corresponds to a specific noise level and can be interpreted as an SNR step of the generative model.
Specifically, the diffusion forward process admits a closed-form SNR expression:
\begin{equation}
\mathrm{SNR}_{\mathrm{DM}}(t)
\triangleq
\frac{\mathbb{E}\!\left[\left\|\sqrt{\bar{\alpha}_t}\,\tmH_0\right\|_2^2\right]}
{\mathbb{E}\!\left[\left\|\sqrt{1-\bar{\alpha}_t}\,\boldsymbol{\eta_t}\right\|_2^2\right]}
=
\frac{\bar{\alpha}_t}{1-\bar{\alpha}_t}.
\end{equation}

Given a pilot-assisted channel observation, its effective SNR can be estimated from the
additive-noise model derived in \eqref{eqn:additivemodel2}. 
We initialize the diffusion reverse process at the step \(t^*\) whose SNR best matches
that of the observation, i.e.,
\begin{equation}
t^{\ast}
=
\arg\min_{t}
\left|
\mathrm{SNR}\!\left(\tilde{\mathbf{Y}}\right)
-
\mathrm{SNR}_{\mathrm{DM}}(t)
\right|.
\end{equation}
Starting the reverse denoising from \(t^*\), the diffusion model iteratively refines
the channel estimate according to
\begin{equation}
\label{DDIM1}
\tmH_{t-1}
=
\mathcal{D}_t\!\left(\tmH_t\right) =
\sqrt{\bar{\alpha}_{t-1}}\,
\mathcal{T}(\tmH_t)
+
\sqrt{1-\bar{\alpha}_{t-1}}\,
\boldsymbol{\varepsilon}_\theta(\tmH_t,t),
\qquad
t = t^{\ast},\dots,1,
\end{equation}
where
\begin{equation}
\mathcal{T}(\tmH_t)
\triangleq
\mathbb{E}[\tmH_0 \mid \tmH_t]
=
\frac{1}{\sqrt{\bar{\alpha}_t}}
\Big(
\tmH_t
-
\sqrt{1-\bar{\alpha}_t}\,
\boldsymbol{\varepsilon}_\theta(\tmH_t,t)
\Big).
\end{equation}
Here, $\mathcal{D}_t(\cdot)$ denotes the diffusion-based denoising operator
implemented by the pretrained diffusion model, where
$\boldsymbol{\varepsilon}_\theta(\tmH_t,t)$ denotes the learned noise predictor.
Note that $\tmH_{t-1}$ is a deterministic function of $\tmH_t$ and $t$ which corresponds to the Denoising Diffusion Implicit Models (DDIM) sampler \cite{song2021ddim}, and $\mathcal{T}(\tilde{\mathbf H}_t)$ denotes the denoised clean-channel estimate,
i.e., the MMSE estimate $\mathbb{E}[\tilde{\mathbf H}_0 \mid \tilde{\mathbf H}_t]$
implicitly used in the DDIM reverse update.
As a result, higher-SNR observations require fewer reverse steps, leading to reduced inference latency compared to initialization from pure Gaussian noise.
$\tmH_0$ is our estimate of the channel in the angular domain and the final estimate of the channel in the spatial domain is the 2-D inverse DFT of $\tmH_0$.
Using \eqref{DDIM1} was introduced in \cite{fesl} and this will serve as our baseline.

\subsection{Likelihood Guidance}
In Section~\ref{sec:pilot}, pilot symbols were leveraged to simplify the original linear observation model into an additive-noise form, enabling diffusion-based denoising for channel estimation.
While the SNR-adaptive reverse process in Section~\ref{sec:snr} aligns the diffusion initialization with the observation noise level, the observation still provides instantaneous sample-level information that can be incorporated during the reverse diffusion dynamics.

To this end, we introduce a likelihood guidance term that enforces first-order consistency between the denoised channel estimate and the pilot-assisted observation.
This guidance exploits the additive-noise observation model to directly incorporate data fidelity into the diffusion-based denoising process.

Since the observation likelihood is defined on the clean channel $\tmH_0$ rather than the noisy diffusion state $\tmH_t$, we incorporate data fidelity using the denoised clean-channel estimate
$\mathcal{T}(\tmH_t) = \mathbb{E}[\tmH_0\mid\tmH_t]$
provided by the diffusion model.

Using this denoised estimate, we approximate the likelihood gradient with respect to the diffusion state as
\begin{equation}
\mathbf{g}_{\mathrm{like}}(\tmH_t)
\triangleq
\nabla_{\tmH_t} \log p(\tmY\mid \tmH_t)
\approx
\frac{1}{\sigma^2}
\Big(
\tmY - \mathcal{T}(\tmH_t)
\Big),
\label{eq:likelihood_grad}
\end{equation}
where $\sigma^2$ denotes the observation noise variance.
Following common practice in diffusion posterior sampling, the Jacobian term of $\mathcal{T}(\tmH_t)$ with respect to $\tmH_t$ is neglected for computational efficiency.

While likelihood guidance provides an effective first-order data consistency mechanism, its reliability depends on the quality of the observation and may degrade in extremely low-SNR regimes.
This issue will be addressed in the joint guidance integration described in the next section.

\subsection{Gram Matrix Guided Diffusion}

Beyond pilot information, additional \emph{second-order} statistical structure can be exploited through the channel Gram matrix which can be estimated from data symbols.
To this end, consider the received symbols in the spatial domain given by
\begin{equation*}
\mathbf{Y}_d = \mathbf{H}\mathbf{X}_d + \mathbf{Z}_d.
\end{equation*}
Forming the sample Gram matrix of \(\mathbf{Y}_d\) yields
\begin{align}
\frac{1}{N_d}
\mathbf{Y}_d \mathbf{Y}_d^{\sf H}
& =
\frac{1}{N_d}
\left(
\mathbf{H}\mathbf{X}_d + \mathbf{Z}_d
\right)
\left(
\mathbf{H}\mathbf{X}_d + \mathbf{Z}_d
\right)^{\sf H} \\
& = \frac{1}{N_d} \mH \mX_d \mX_d ^{\sf H}\mH^{\sf H} + \frac{1}{N_d} \mZ_d \mZ_d^{\sf H} + 
\frac{1}{N_d} \mH \mX_d \mZ_d^{\sf H} + 
\frac{1}{N_d} \mZ_d \mX_d^{\sf H} \mH^{\sf H}.
\end{align}
The matrix product $\frac{1}{N_d} \mX_d \mX_d^{\sf H}$ can be written as 
$\frac{1}{N_d} \sum_i \mathbf{X}_{d,i} \mathbf{X}_{d,i}^{\sf H}$ where $\mathbf{X}_{d,i}$ is the $i$th column of the matrix $\mX_d$.
From the weak law of large numbers, we can see that  $\frac{1}{N_d} \mX_d \mX_d^{\sf H} \xrightarrow{N_d \rightarrow \infty} \mI$ since the columns of $\mX_d$ are uncorrelated data symbols \footnote{We assume appropriate interleaving if there is any error correction coding}.
Assuming sufficiently long data blocks, the cross terms vanish after averaging, leading to
\begin{equation}
\frac{1}{N_d}
\mathbf{Y}_d \mathbf{Y}_d^{\sf H}
\approx
\mathbf{H}\mathbf{H}^{\sf H}
+
\sigma_d^2 \mathbf{I}.
\end{equation}
Then an estimate of the channel Gram matrix in the spatial domain can be obtained according to
\begin{equation}
\hat{\mR} = \widehat{\mathbf{H}\mathbf{H}^{\sf H}}
=
\mathcal{P}\left( \frac{1}{N_d}
\mathbf{Y}_d \mathbf{Y}_d^{\sf H}
-
\sigma_d^2 \mathbf{I}\right),
\end{equation}
Where $\mathcal{P}$ is a projection operator that is designed to robustify the estimate; for example, by enforcing positive semidefiniteness of $\hat{\mR}$. We also consider other shrinkage operators explained later.
An estimate of the Gram matrix in the angular domain is then obtained using 
\begin{equation}
    \hat{\tmR} = \Phi_{N_R} \hat{\mR} \Phi_{N_R}^{\sf T}.
\end{equation}

We incorporate this estimated gram matrix as a guidance term during the reverse diffusion process by penalizing the mismatch between the gram matrix implied by the current diffusion state and the estimated channel covariance. Specifically, we define the gram matrix-guided objective
\begin{equation}
f(\tmH_t)
=
-
\left\|
\tmH_t \tmH_t^{\sf H}
-
\hat{\tmR}
\right\|_{\sf F}^2 .
\end{equation}
Its gradient yields the gram matrix guidance direction
\begin{equation}
\mathbf{g}_{\text{Gram}}(\tmH_t)
\;\triangleq\;
\nabla_{\tmH_t} f(\tmH_t)
=
4\big(\widehat{\tilde{\mathbf{R}}} - \tmH_t \tmH_t^{\sf H}\big)\tmH_t.
\end{equation}

Finally, we jointly inject the likelihood-based and Gram-based guidance into the diffusion denoising process by modifying the reverse update as
\begin{equation}
\tmH_{t-1}
=
\mathcal{D}_t(\tmH_t)
+
\lambda_{\text{like},t}\,
\mathbf g_{\text{like}}(\tmH_t)
+
\lambda_{\text{Gram},t}\,
\mathbf g_{\text{Gram}}(\tmH_t),
\qquad
t = t^{\ast},\dots,1 ,
\label{eq:joint_update}
\end{equation}
where $\mathcal{D}_t(\cdot)$ denotes the pretrained diffusion denoiser at step $t$.
The overall procedure is summarized in Table~\ref{alg:snr_cov_diffusion}.

\begin{algorithm}[ht]
\caption{Diffusion with Gram Matrix Guidance}
\label{alg:snr_cov_diffusion}
\begin{algorithmic}[1]
\Require
Pilot observation $\mathbf{Y}_p$, data observation $\mathbf{Y}_d$,
pilot matrix $\mathbf{X}_p$,
pretrained diffusion denoiser $\{\mathcal{D}_t\}_{t=1}^T$
\Ensure
Channel estimate $\widehat{\mathbf{H}}$

\State $\tilde{\mathbf{Y}} \leftarrow \Phi_{N_R} \mathbf{Y}_p \mathbf{X}_p^{\sf H}
(\mX_p \mX_p^{\sf H})^{-1} \Phi_{N_T}^{\sf T}$
\Comment{Pilot decorrelation in angular domain}

\State $t^\ast \leftarrow \arg\min_t \big| \mathrm{SNR}(\tilde{\mathbf{Y}}) - \mathrm{SNR}_{\mathrm{DM}}(t) \big|$

\State $\tmH_{t^\ast} \leftarrow (1+\sigma^2)^{-1/2}\,\tilde{\mathbf{Y}}$
\Comment{Variance normalization}

\State $\widehat{\mathbf{R}} \leftarrow \mathcal{P}
\left(\frac{1}{N_d}\mathbf{Y}_d \mathbf{Y}_d^{\sf H} - \sigma_d^2 \mathbf{I}\right)$
\Comment{Sample Gram matrix in the spatial domain}

\State $\widehat{\tilde{\mathbf{R}}} \leftarrow \Phi_{N_R} \widehat{\mathbf{R}} \Phi_{N_T}^{\sf H}$
\Comment{Estimate of the Gram matrix in the angular domain}

\For{$t = t^\ast$ \textbf{to} $1$}
    \State $\mathbf{g}_{\text{Gram}} \leftarrow 4\big(\widehat{\tilde{\mathbf{R}}} - \tmH_t \tmH_t^{\sf H}\big)\tmH_t$
    \Comment{Guidance based on the Gram matrix estimate}
    \State $\mathbf{g}_{\text{like}} \leftarrow \frac{1}{\sigma^2}\big(\tilde{\mathbf{Y}} - \mathcal{T}(\tmH_t) \big)$
    \Comment{Likelihood guidance based on Tweedie estimate $\mathcal{T}(\tmH_t)$}
    
    \State $\tmH_{t-1} \leftarrow \mathcal{D}_t(\tmH_t)
    + \lambda_{\text{Gram},t}\,\mathbf{g}_{\text{Gram}}
    + \lambda_{\text{like},t}\,\mathbf{g}_{\text{like}}$
\EndFor

\State $\widehat{\mathbf{H}} \leftarrow 
\Phi_{N_R}^{\sf H} \tmH_0 \Phi_{N_T}^*$
\Comment{Mapping back to spatial domain through 2-D IDFT operation}

\end{algorithmic}
\end{algorithm}

To match the intrinsic step size of the reverse process, we adopt a diffusion-aware scaling for the likelihood strength, proportional to the noise schedule $\beta_t$,
\begin{equation}
\lambda_{\text{like},t} \propto \beta_t ,
\end{equation}
which serves as a baseline step-size normalization.
The effective likelihood influence is further modulated by the SNR-aware gating mechanism described below.

When likelihood guidance is incorporated into the reverse diffusion dynamics, its reliability may depend on both the observation SNR and the underlying channel statistics.
In particular, for certain channel families, naive likelihood guidance can become unstable in the very-low SNR regime and introduce excessive sample-level perturbations that degrade the denoising performance.
To address this issue, we introduce an SNR-aware gating mechanism as a robustness safeguard, which modulates the effective strength of likelihood guidance based on the estimated observation SNR.
This gating mechanism is beneficial for channel families where naive likelihood guidance becomes unreliable at very-low SNRs (e.g., QuaDRiGa LOS), while remaining neutral for channels where likelihood guidance is inherently stable (e.g., 3GPP). We have
\begin{equation}
\lambda_{\text{like},t}
=
\lambda_{\text{like}}
\, \beta_t \,
w\!\left(\mathrm{SNR}(\tilde{\mathbf{Y}})\right),
\label{eq:like_scaling_gated}
\end{equation}
where $\lambda_{\text{like}}$ is a scalar, and the gating function $w(\cdot)$ is defined as
\begin{equation}
w\!\left(\mathrm{SNR}(\tilde{\mathbf{Y}})\right)
=
\frac{1}{1+\exp\!\left(-\frac{\mathrm{SNR}(\tilde{\mathbf{Y}})-\mathrm{SNR}_0}{\Delta}\right)}.
\label{eq:snr_gate}
\end{equation}
Here, $\mathrm{SNR}_0$ controls the transition point at which likelihood guidance becomes effective, and $\Delta$ determines the smoothness of the transition.
The gating function depends solely on the observation SNR and is applied uniformly across reverse diffusion steps, thereby decoupling likelihood reliability from the diffusion time index.

In contrast, the gram matrix guidance enforces a global, second-order structural constraint on the channel estimate.
Since it operates at a different level from the stochastic reverse update, its influence does not need to strictly follow the diffusion step size.
In practice, we adopt a sublinear scaling
\begin{equation}
\lambda_{\text{Gram},t} \propto \sqrt{\beta_t},
\label{eq:cov_scaling}
\end{equation}
which empirically provides effective structural regularization without overwhelming the reverse diffusion dynamics. Importantly, the overall magnitude of Gram guidance is controlled by a global scaling factor embedded in $\lambda_{\text{Gram},t}$, which provides a mechanism to trade off structural regularization strength against Gram estimation uncertainty. This degree of freedom becomes critical in coherence-time limited regimes and will be exploited in Sec.~\ref{sec:robust}.
To ensure numerical stability, the Gram matrix guidance update is further restricted using a norm-based clipping strategy; see Appendix~\ref{app:impl}, paragraph \emph{Stability of Gram Guidance}.

\section{Simulation Results}
\label{sec:sim}

\subsection{Channel Models, Baselines, and Evaluation Setup}
\label{sec:setup}
We consider a massive MIMO channel estimation scenario with $(N_r, N_t) = (64, 16)$, consistent with the experimental configuration in~\cite{fesl}.
We evaluate the proposed diffusion-based channel estimation framework with likelihood and Gram matrix guidance on two widely used datasets following~\cite{fesl}: a 3GPP spatial channel model and the QuaDRiGa channel simulator (v2.4).

\paragraph{Structural differences of the induced channel statistics.}
Although both pipelines provide access to ground-truth channel realizations $\mathbf{H}$ for evaluation, the induced spatial second-order structure depends on the scenario and simulator settings.
In our experiments, we explicitly analyze the eigenvalue distribution of the realization-level Gram matrix $\mathbf{H}\mathbf{H}^{\sf H}$.
We observe that the 3GPP setting considered here yields a more distributed eigen-spectrum (higher spectral entropy), whereas the QuaDRiGa setting yields a more concentrated eigen-spectrum (lower spectral entropy) with larger variability across realizations.
We use this empirical characterization only to contextualize the trends observed in later results.

\paragraph{Estimators and baselines.}
We compare the following estimators.
\begin{itemize}
\item \textbf{DM (primary baseline):} the SNR-matched deterministic diffusion estimator without explicit guidance; all guided variants are compared against this baseline.
\item \textbf{DM + Likelihood:} diffusion with first-order (sample-level) likelihood guidance.
\item \textbf{DM + Gram:} diffusion with second-order Gram guidance only.
\item \textbf{DM + Gram + Likelihood:} diffusion with joint first- and second-order guidance.
Two Gram modes are considered:
(i) oracle Gram matrix $\mathbf{H}\mathbf{H}^{\sf H}$,
and (ii) estimated Gram matrix obtained from data symbols.
\item \textbf{Genie-LMMSE (3GPP only):} genie-aided linear minimum-mean squared error estimator.
\end{itemize}
The 3GPP channel model corresponds to a mixture jointly-Gaussian model where the covariances of the components of the mixture are available. 
This enables computing a genie-aided bound using a linear minimum-mean squared error estimator, where the genie reveals the covariance of the component from which the channel realization is drawn.
In contrast, for the QuaDRiGa pipeline, channel realizations are generated from a highly complex, environment-dependent stochastic geometry model.
As a result, the underlying channel distribution does not admit a tractable or stationary second-order statistical characterization.
Consequently, an oracle covariance required by Genie-LMMSE is not well-defined in this setting.

\paragraph{Evaluation setup.}
All methods are evaluated over $\mathrm{SNR}\in[-15,5]$~dB.
Performance is measured by the normalized mean-squared error (NMSE) in the spatial channel domain,
\begin{equation}
\mathrm{NMSE}_{\mathrm{ch}}
\triangleq
\mathbb{E}\!\left[
\frac{\|\mathbf H-\widehat{\mathbf H}\|_{\sf F}^2}
{\|\mathbf H\|_{\sf F}^2}
\right],
\end{equation}
where $\widehat{\mathbf{H}}$ is obtained by applying the inverse DFT to the final angular-domain estimate produced by the diffusion model, as described in Algorithm~\ref{alg:snr_cov_diffusion}.
The expectation is taken over channel realizations and noise.

Unless otherwise specified, we use $N_d=2000$ data symbols for estimating the instantaneous Gram matrix from data observations.
As will be shown, once $N_d$ exceeds the stability threshold of the Gram-guided reverse diffusion dynamics (around a few hundred samples in our 3GPP setting), further increasing $N_d$ has negligible impact on the end-to-end NMSE.

\subsection{Results on 3GPP}

\begin{figure}[t]
  \centering
    \begin{tikzpicture}
        \begin{axis}[
          width=12.5cm,height=8.3cm,
          xmin=-15,xmax=5,
          ymin=4e-2,ymax=1,
          ymode=log,
          axis lines=left,
          xlabel={SNR [dB]},
          ylabel={NMSE},
          label style={font=\small},
          tick label style={font=\small},
          tick align=outside,
          tick style={line width=0.4pt, black!60},
          xtick={-15,-13,...,5}, 
          minor xtick={-15,-14,...,5},
          grid=both,
          major grid style={line width=0.2pt, draw=black!15},
          minor grid style={line width=0.15pt, draw=black!8},
          legend style={
            at={(0.02,0.02)}, 
            anchor=south west,
            draw=black!30,
            fill=white,
            fill opacity=0.95,
            text opacity=1,
            font=\small,
            rounded corners=1pt,
            inner sep=2pt,
          },
          legend cell align=left,
        ]
        
        \addplot+[color=cDM, line width=2pt, mark=o, mark size=2.6pt, mark repeat=2]
        coordinates {
        (-15,0.956674993) (-14,0.932828784) (-13,0.898368776) (-12,0.851259053) (-11,0.79441011)
        (-10,0.727668762) (-9,0.654289067) (-8,0.574952424) (-7,0.499485046) (-6,0.427704483)
        (-5,0.364863485) (-4,0.309671074) (-3,0.262683958) (-2,0.220814109) (-1,0.1872347)
        (0,0.157922253) (1,0.133347675) (2,0.112339757) (3,0.095159866) (4,0.079534099)
        (5,0.067043602)
        };
        \addlegendentry{DM}

        \addplot+[color=cLIKE, line width=2pt, mark=o, mark size=2.6pt, mark repeat=2, dashed]
        coordinates {
        (-15,0.9525881) (-14,0.9248568) (-13,0.8880149) (-12,0.8341098) (-11,0.7732138)
        (-10,0.7018799) (-9,0.6236374) (-8,0.5442631) (-7,0.4674852) (-6,0.3998596)
        (-5,0.3414079) (-4,0.2907076) (-3,0.2473322) (-2,0.2094728) (-1,0.1783003)
        (0,0.1516210) (1,0.1287495) (2,0.1088099) (3,0.09236980) (4,0.07785374)
        (5,0.06560829)
        };
        \addlegendentry{DM + Likelihood}
        
        \addplot+[color=cGRAM, line width=2pt, mark=s, mark size=2.6pt, mark repeat=2]
        coordinates {
        (-15,0.7998869) (-14,0.7224673) (-13,0.6470816) (-12,0.5771204) (-11,0.5121256)
        (-10,0.4515603) (-9,0.3962944) (-8,0.3437672) (-7,0.2971370) (-6,0.2554634)
        (-5,0.2186876) (-4,0.1864926) (-3,0.1583882) (-2,0.1333226) (-1,0.1125727)
        (0,0.09487968) (1,0.07938926) (2,0.06656365) (3,0.05602067) (4,0.04675360)
        (5,0.03917487)
        };
        \addlegendentry{DM + Gram}

        \addplot+[color=cORACLE, line width=2pt, mark=triangle*, mark size=2.8pt, mark repeat=2]
        coordinates {
        (-15,0.7991148) (-14,0.7217503) (-13,0.6479988) (-12,0.5772156) (-11,0.5121858)
        (-10,0.4507651) (-9,0.3942465) (-8,0.3425176) (-7,0.2951238) (-6,0.2532761)
        (-5,0.2170824) (-4,0.1842092) (-3,0.1562552) (-2,0.1317736) (-1,0.1111409)
        (0,0.09336518) (1,0.07842468) (2,0.06561569) (3,0.05512220) (4,0.04600573)
        (5,0.03846021)
        };
        \addlegendentry{DM + Gram (oracle) + Likelihood}
        
        \addplot+[color=cEST, line width=2pt, mark=diamond*, mark size=2.6pt, mark repeat=2]
        coordinates {
        (-15,0.8046212) (-14,0.7268641) (-13,0.6512193) (-12,0.5809162) (-11,0.5137955)
        (-10,0.4520712) (-9,0.3944145) (-8,0.3428931) (-7,0.2966368) (-6,0.2543182)
        (-5,0.2176813) (-4,0.1852136) (-3,0.1575337) (-2,0.1325100) (-1,0.1118462)
        (0,0.09405626) (1,0.07904492) (2,0.06623647) (3,0.05558591) (4,0.04664312)
        (5,0.03911159)
        };
        \addlegendentry{DM + Gram (est) + Likelihood}
        
        \addplot+[color=cLMMSE, line width=2pt, mark=x, mark size=3.0pt, mark repeat=2]
        coordinates {
        (-15,0.654871892) (-14,0.614210594) (-13,0.571697074) (-12,0.528890459) (-11,0.485627649)
        (-10,0.44325059) (-9,0.402287903) (-8,0.361847727) (-7,0.323922238) (-6,0.28793767)
        (-5,0.25426745) (-4,0.223241458) (-3,0.194940835) (-2,0.169181755) (-1,0.145924788)
        (0,0.125577844) (1,0.10756921) (2,0.091593267) (3,0.077720248) (4,0.06574859)
        (5,0.055502986)
        };
        \addlegendentry{Genie-LMMSE}
        
        \end{axis}
        \end{tikzpicture}
  \caption{3GPP: NMSE vs. SNR.}
  \label{fig:3gpp_nmse}
\end{figure}
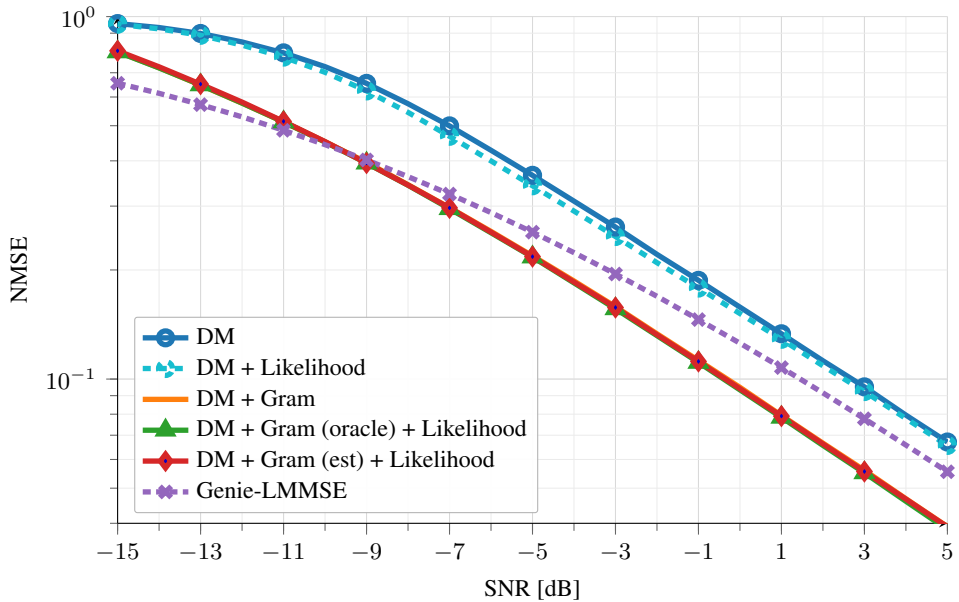

Fig.~\ref{fig:3gpp_nmse} reports NMSE versus SNR for the 3GPP case.
We observe consistent performance differences among the compared methods.
First, the SNR-matched deterministic diffusion estimator provides a strong baseline, which we attribute to its initialization that aligns the reverse diffusion process with the observation noise level.
Using Gram matrix guidance provides a substantial gain by injecting additional second-order structural information estimated from data symbols.

In the 3GPP channel model case, incorporating likelihood guidance alone does not yield a significant gain. This suggests that, under the considered 3GPP setting, the SNR-matched diffusion prior is already relatively well-aligned with the data, leaving limited room for additional correction from likelihood guidance alone; we revisit this contrast in the QuaDRiGa results.

To quantify the gain, DM + Gram + Likelihood yields an SNR advantage of approximately $5$--$7$~dB at representative NMSE operating points (e.g., around $\mathrm{NMSE}\approx 2\times10^{-1}$), compared to the unguided DM baseline.
Equivalently, at a fixed SNR, it reduces NMSE across the entire SNR range, with the most pronounced improvement appearing in the low-to-moderate SNR regime where pilot-only (first-order) constraints are weaker.

We include a Genie-LMMSE curve in Fig.~\ref{fig:3gpp_nmse} as an oracle reference that conditions on realization-specific covariance information, and is therefore unattainable for estimators relying only on first-order observations.
However, incorporating Gram-matrix guidance enables the diffusion estimator to leverage a realization-level second-order statistic $\mathbf{H}\mathbf{H}^{\sf H}$ (or its data-driven estimate), which empirically closes the gap to the Genie-LMMSE reference and outperforms it over a range of moderate-to-high SNRs in our setup.

We emphasize that Genie-LMMSE and our Gram-guided diffusion condition on different forms of side information -- a model-level joint distribution prior versus a realization-level second-order statistic.
Since these information sets are not nested, neither method is guaranteed to uniformly dominate the other; consequently, the relative ordering may vary across SNR, and in our experiments the Gram-guided diffusion is possible to match or outperform the Genie-LMMSE reference in the moderate-to-high SNR regime.

With $N_d=2000$, the curves corresponding to oracle and estimated Gram matrix guidance are nearly indistinguishable in
Fig.~\ref{fig:3gpp_nmse}.
This indicates that once the Gram matrix estimation error falls below the stability threshold, the reverse diffusion dynamics become insensitive to residual estimation perturbations, consistent with the threshold behavior discussed in Sec.~\ref{sec:robust}.
In practice, $N_d$ can be reduced substantially as long as it remains above the threshold regime—with minimal impact on the end-to-end NMSE.

\subsection{Results on QuaDRiGa}

\begin{figure}[t]
  \centering
    \begin{tikzpicture}
        \begin{axis}[
          width=12.5cm,height=8.3cm,
          xmin=-15,xmax=5,
          ymin=4e-2,ymax=1,
          ymode=log,
          axis lines=left,
          xlabel={SNR [dB]},
          ylabel={NMSE},
          label style={font=\small},
          tick label style={font=\small},
          tick align=outside,
          tick style={line width=0.4pt, black!60},
          xtick={-15,-13,...,5}, 
          minor xtick={-15,-14,...,5},
          grid=both,
          major grid style={line width=0.2pt, draw=black!15},
          minor grid style={line width=0.15pt, draw=black!8},
          legend style={
            at={(0.98,0.98)}, 
            anchor=north east,
            draw=black!30,
            fill=white,
            fill opacity=0.95,
            text opacity=1,
            font=\small,
            rounded corners=1pt,
            inner sep=2pt,
          },
          legend cell align=left,
        ]

        \addplot+[color=cDM, line width=2pt, mark=o, mark size=2.6pt, mark repeat=2]
        coordinates {
        (-15,0.345148116) (-14,0.28891772) (-13,0.251942813) (-12,0.224093035) (-11,0.200874627)
        (-10,0.177105069) (-9,0.162511826) (-8,0.150072262) (-7,0.142880231) (-6,0.133797809)
        (-5,0.126413807) (-4,0.119556457) (-3,0.112782106) (-2,0.106160231) (-1,0.098546132)
        (0,0.090614289) (1,0.082667232) (2,0.074467674) (3,0.067861624) (4,0.059568506)
        (5,0.053351436)
        };
        \addlegendentry{DM}

        \addplot+[color=cLIKE, line width=2pt, mark=o, mark size=2.6pt, mark repeat=2, dashed]
        coordinates {
        (-15,0.3120825) (-14,0.2637613) (-13,0.2285920) (-12,0.2030980) (-11,0.1810158)
        (-10,0.1627318) (-9,0.1495765) (-8,0.1386342) (-7,0.1309456) (-6,0.1229533)
        (-5,0.1156248) (-4,0.1082929) (-3,0.1007664) (-2,0.09259009) (-1,0.08497594)
        (0,0.07772791) (1,0.07067811) (2,0.06403799) (3,0.05759114) (4,0.05179929)
        (5,0.04622400)
        };
        \addlegendentry{DM + Likelihood}
        
        \addplot+[color=cGRAM, line width=2pt, mark=s, mark size=2.6pt, mark repeat=2]
        coordinates {
        (-15,0.1748635) (-14,0.1538816) (-13,0.1387912) (-12,0.1285072) (-11,0.1205567)
        (-10,0.1153305) (-9,0.1111535) (-8,0.1070588) (-7,0.1037800) (-6,0.1005809)
        (-5,0.09726343) (-4,0.09411472) (-3,0.09060667) (-2,0.08586457) (-1,0.08137912)
        (0,0.07657793) (1,0.07111976) (2,0.06517617) (3,0.06066789) (4,0.05418116)
        (5,0.04941193)
        };
        \addlegendentry{DM + GRAM}
        
        \addplot+[color=cORACLE, line width=2pt, mark=triangle*, mark size=2.8pt, mark repeat=2]
        coordinates {
        (-15,0.1764893) (-14,0.1529106) (-13,0.1383094) (-12,0.1280172) (-11,0.1203346)
        (-10,0.1143455) (-9,0.1093437) (-8,0.1048839) (-7,0.1006874) (-6,0.09638820)
        (-5,0.09208596) (-4,0.08760896) (-3,0.08263041) (-2,0.07669161) (-1,0.07099428)
        (0,0.06529442) (1,0.05927842) (2,0.05366533) (3,0.04882218) (4,0.04388248)
        (5,0.03951930)
        };
        \addlegendentry{DM + GRAM (oracle) + Likelihood}
        
        \addplot+[color=cEST, line width=2pt, mark=diamond*, mark size=2.6pt, mark repeat=2]
        coordinates {
        (-15,0.1759127) (-14,0.1538754) (-13,0.1392316) (-12,0.1280777) (-11,0.1205633)
        (-10,0.1145857) (-9,0.1096064) (-8,0.1048951) (-7,0.1004672) (-6,0.09649553)
        (-5,0.09214246) (-4,0.08755556) (-3,0.08261013) (-2,0.07682554) (-1,0.07120566)
        (0,0.06516127) (1,0.05936537) (2,0.05382114) (3,0.04885961) (4,0.04394368)
        (5,0.03962712)
        };
        \addlegendentry{DM + GRAM (est) + Likelihood}        
        
        \end{axis}
        \end{tikzpicture}
  \caption{QuaDRiGa: NMSE vs. SNR.}
  \label{fig:quadriga_nmse}
\end{figure}
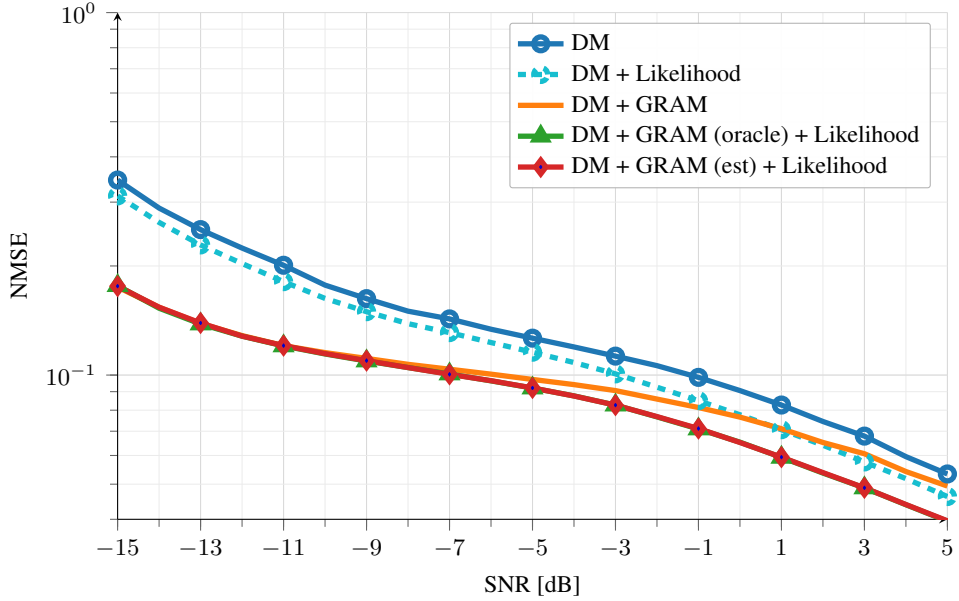

Fig.~\ref{fig:quadriga_nmse} shows the corresponding results under the QuaDRiGa channel model.
Incorporating Gram matrix guidance provides a substantial improvement over the unguided diffusion baseline.
In particular, DM + Gram + Likelihood achieves a comparable $5$--$7$~dB SNR advantage at representative operating points, while the contribution of likelihood guidance is notably more pronounced than in the 3GPP case.

The distinct second-order structures observed in Sec.~\ref{sec:setup} provide a useful lens to interpret the results.
Under the considered QuaDRiGa setting, the eigen-spectrum of the realization-level Gram matrix $\mathbf{H}\mathbf{H}^{\sf H}$ is more concentrated and exhibits larger variability across realizations.
A concentrated spectrum indicates that most channel energy is captured by a few dominant spatial eigen-modes.
As a result, Gram guidance mainly constrains a low-dimensional energy subspace and the associated energy allocation, while leaving residual ambiguity within that subspace.
Many different channel realizations can share a similar Gram matrix, since there exist transformations that preserve $\mathbf{H}\mathbf{H}^{\sf H}$.
Therefore, Gram guidance may reduce ambiguity but it may not fully determine the realization.

In addition to yielding a noticeable improvement over the unguided diffusion baseline, likelihood guidance further enhances the performance when combined with Gram matrix guidance by enforcing sample-level data consistency and selecting realizations that better match the received observation.
This effect is most visible at moderate-to-high SNR, where the observation model becomes more informative.

Importantly, the QuaDRiGa results illustrate a practically relevant regime where Gram-matrix guidance can be leveraged without an explicit parametric statistical model for the channel and without oracle covariance information.
The receiver can still form a data-driven estimate of a realization-level second-order statistic from the received data symbols $\mathbf{Y}_d$.
This enables the diffusion process to exploit structural information that is difficult to recover reliably from purely pilot-driven, first-order methods, without requiring additional supervision or access to ground-truth channels.
Consequently, the proposed method remains applicable to realistic measurement pipelines where only received signals are available.

Finally, with $N_d=2000$, the performance is already in the stable regime: further increasing $N_d$ yields negligible additional NMSE improvement.

\subsection{Complexity Analysis}
We compare the online computational cost of three estimators: (i) the SNR-matched deterministic diffusion baseline, (ii) the proposed Gram-guided diffusion estimator, and (iii) the genie-aidedLMMSE benchmark. Throughout, the channel in the angular domain is represented as $\tilde{\mathbf{H}}\in\mathbb{C}^{N_R\times N_T}$, with vectorized dimension $N \triangleq N_RN_T$. We denote computational cost by $\mathcal{K}(\cdot)$ to avoid confusion with covariance matrices.

\paragraph{Genie-LMMSE.}
The genie-aided LMMSE baseline operates on the vectorized channel of dimension
$N$ and requires solving an $N\times N$ linear system or equivalently inverting an $N\times N$ matrix per realization.
A standard factorization therefore incurs $\mathcal{O}(N^3)$ complexity, i.e.,
\begin{equation}
\mathcal{K}_{\text{Genie}}=\mathcal{O}(N^3).
\end{equation}

\paragraph{SNR-matched DM.}
The DM baseline avoids large matrix inversion. Its online cost consists of: (i) transforming pilot-based quantities to the angular domain and back, and (ii) evaluating the denoiser for a truncated number of reverse steps enabled by SNR-matched initialization. Specifically, the reverse process starts from $t^\ast$ (rather than running all $T$ steps), so the number of denoiser evaluations scales with $t^\ast$, which decreases with increasing SNR. Let $\mathcal{K}_{\text{NN}}$ denote the cost of a single denoiser forward pass; since our denoiser architecture is identical to the DM reference, $\mathcal{K}_{\text{NN}}$ is unchanged between DM and GRAM-DIFF. The DM complexity is therefore
\begin{equation}
\mathcal{K}_{\text{DM}}
=
\underbrace{\mathcal{O}(N\log N)}_{\text{FFT transforms}}
+
\underbrace{t^\ast\,\mathcal{K}_{\text{NN}}}_{\text{denoiser evaluations}}
+
\underbrace{\mathcal{O}(N\log N)}_{\text{IFFT transforms}}
=
\mathcal{O}(N\log N)+t^\ast\mathcal{K}_{\text{NN}}.
\end{equation}
Multiplication by $\Phi_{N_R}$ and $\Phi_{N_T}$ correspond to 1-D DFT operations and hence, can be implemented using fast Fourier transforms.
A concrete FLOPs count for $\mathcal{K}_{\text{NN}}$ under the CNN configuration is provided in the Appendix.

\paragraph{GRAM-DIFF.}
Relative to DM, GRAM-DIFF adds two lightweight components: likelihood guidance and Gram guidance. The likelihood guidance reuses the denoiser output via a denoised estimate such as $\mathcal{T}(\tilde{\mathbf{H}}_t)$ and introduces only elementwise operations, contributing $\mathcal{O}(N)$ per reverse step. The Gram guidance update involves matrix multiplications of the form $(\widehat{\tilde{\mathbf{R}}}-\tilde{\mathbf{H}}_t\tilde{\mathbf{H}}_t^{\mathrm{H}})\tilde{\mathbf{H}}_t$, whose per-step cost scales as $\mathcal{O}(N_R^2N_T)$. In addition, GRAM-DIFF estimates the Gram term from data symbols, e.g., $\hat{\mathbf{R}} \propto \mathbf{Y}_d\mathbf{Y}_d^{\mathrm{H}}$, which incurs a one-time cost $\mathcal{O}(N_R^2N_d)$ for $\mathbf{Y}_d\in\mathbb{C}^{N_R\times N_d}$. Overall,
\begin{equation}
\mathcal{K}_{\text{GRAM-DIFF}}
=
\mathcal{K}_{\text{DM}}
+
\underbrace{\mathcal{O}(N_R^2N_d)}_{\text{one-time Gram estimation}}
+
t^\ast\Big(
\underbrace{\mathcal{O}(N)}_{\text{likelihood guidance}}
+
\underbrace{\mathcal{O}(N_R^2N_T)}_{\text{Gram guidance}}
\Big).
\end{equation}
Since the added guidance operations do not introduce any $\mathcal{O}(N^3)$ term, GRAM-DIFF remains substantially cheaper than Genie-LMMSE while improving NMSE over the unguided diffusion baseline.

\begin{table}[t]
\centering
\caption{Online computational complexity comparison per realization.}
\label{tab:complexity_comparison}
\begin{tabular}{l l}
\toprule
Estimator & Online complexity $\mathcal{K}(\cdot)$ \\
\midrule
SNR-matched DM &
$\mathcal{O}(N\log N) + t^\ast\,\mathcal{K}_{\text{NN}}$ \\[2pt]
GRAM-DIFF &
$\mathcal{O}(N\log N) + t^\ast\,\mathcal{K}_{\text{NN}}
+ \mathcal{O}(N_R^2N_d)
+ t^\ast\!\left(\mathcal{O}(N)+\mathcal{O}(N_R^2N_T)\right)$ \\[2pt]
Genie-LMMSE &
$\mathcal{O}(N^3)$ \\
\bottomrule
\end{tabular}
\end{table}

\subsection{Remarks}
We emphasize that our focus is on a hybrid solution whose performance gracefully degrades with the quality of estimate of $\mR$. 
We have focused on using $\hat{\mR}$ as additional side information and designed a channel estimation algorithm that defaults to a likelihood-guided DM built upon the SNR-matched DM algorithm in \cite{fesl} when $\hat{\mR}$ is not available.
The computational price to pay for this is that the complexity per denoising step of the proposed approach depends on $N_R$ as $\mathcal{O}(N_R^2)$.

If our interest is only in the case when perfect knowledge of $\mR$ (and hence $\tmR$) is available at the receiver, 
we may be able to decrease the computational complexity by using the following approach (similar in spirit to that in \cite{jagannatham2006whitening}).
The channel matrix $\tmH$ can be decomposed using a singular-value decomposition as 
\[
\tmH = \tmU \tmsigma \tmV^{\sf H},
\]
where the columns of $\tmU$ denote the eigenvectors of $\tmR$ and the diagonal elements of $\tmsigma$ denote the eigenvalues of $\tmR$.
We can transform the channel realizations in $\mathcal{H}$ to a dataset of realizations of $\tmV$.
We can train a conditional diffusion model to learn to generate $\tmV$ conditioned on $\tmU$ and $\tmsigma$.
If $\tmV$ is independent of $\tmU$ and $\tmsigma$, then 
the conditioning is not required and our diffusion model can be trained to sample $\tmV$ based on the dataset.
During inference, we can preprocess $\tmY$ by premultiplying by $\tmsigma^{-1} \tmU^{\sf H}$
and solving only for $\tmV$. 
Since $\tmV \in \mathbb{C}^{N_T \times N_T}$, when $N_T < N_R$, this has lower computational complexity per denoising step than the proposed pipeline.
However, this requires a one-time eigenvalue decomposition with complexity $\mathcal{O}(N_R^3)$.
The noise variance per dimension will also be different due to the premultiplication by $\tmsigma^{-1}$.
This architecture will be studied in future work.

\subsection{Coherence-Time Robustness of GRAM-DIFF}
\label{sec:robust}

Recall that in Algorithm~\ref{alg:snr_cov_diffusion} the data block length $N_d$ (i.e., the number of data symbols within one channel coherence interval) affects the diffusion process through the accuracy of the estimated Gram matrix $\widehat{\mathbf R}$ used in the Gram matrix guidance term. 
In the preceding sections, we primarily focused on regimes where sufficiently many data symbols are available such that the Gram estimate is reliable and the guidance strength can be applied without stability concerns. 
In practical systems, however, coherence time, latency constraints, and signaling overhead often limit the number of usable data symbols. 
This motivates a careful examination of the robustness of Gram-guided diffusion in coherence-time limited regimes and the identification of operating conditions under which performance gains can be maintained.

A key observation of this work is that Gram-guided diffusion can remain effective even under severe sample limitations by appropriately adjusting the Gram guidance strength.
Rather than treating the estimated Gram matrix as a hard constraint, the algorithm injects Gram information as a soft structural bias whose influence can be continuously controlled.
From a system perspective, maintaining robust performance under limited data availability is therefore more critical than marginal performance gains in regimes where $N_d$ is already large.

\vspace{0.5em}
\noindent
\textbf{Estimation Error Scaling and Stability Interpretation.}
We quantify the Gram estimation accuracy using the normalized mean-square error
\begin{equation}
\mathrm{NMSE}_{\mathrm{R}}
\triangleq
\mathbb{E}\!\left[
\frac{\|\widehat{\mathbf R}-\mathbf R\|_{\sf F}^2}
{\|\mathbf R\|_{\sf F}^2}
\right].
\end{equation}
For sample Gram matrix estimation, this error approximately follows the scaling law
\begin{equation}
\mathrm{NMSE}_{\mathrm{R}}
=
\mathcal{O}\!\left(
\frac{C(\mathrm{SNR})}{N_d}
\right),
\end{equation}
where $C(\mathrm{SNR})$ is a SNR-dependent coefficient capturing the effective noise level.

When Gram guidance is applied with a fixed aggressive scaling, this estimation error does not translate into a smooth performance degradation.
Instead, we observe a pronounced threshold behavior:
below a critical error level, the reverse diffusion trajectory remains stable and the end-to-end channel NMSE saturates to a performance plateau, whereas exceeding this threshold causes erroneous Gram guidance to dominate the update direction and degrade performance.

Importantly, this threshold is not solely determined by the Gram estimation error itself, but also depends on the effective magnitude of the guidance injection controlled by the Gram guidance scaling in Algorithm~\ref{alg:snr_cov_diffusion}.
By reducing the guidance strength, the stability region of the reverse diffusion dynamics can be substantially expanded, allowing the algorithm to tolerate significantly noisier Gram estimates.

\vspace{0.5em}
\noindent
\textbf{Feasible Data Block Length Under Adaptive Guidance.}
Based on the above observations, practical feasibility is better interpreted in terms of a controllable operating regime rather than a hard minimum data length requirement.
Specifically, Gram guidance is considered beneficial if
\begin{equation}
\mathrm{NMSE}_{\mathrm{ch}}^{\mathrm{GRAM-DIFF}}(\mathrm{SNR},N_d)
<
\mathrm{NMSE}_{\mathrm{ch}}^{\mathrm{DM+Lik}}(\mathrm{SNR}),
\quad \forall \; \mathrm{SNR} \in \mathcal{S},
\label{eq:nd_feasible}
\end{equation}
where $\mathcal{S}$ denotes the target SNR operating range.

When a fixed guidance strength is used without adaptation, this condition induces an apparent lower bound on $N_d$, which is approximately $N_d\approx200$ for the 3GPP scenario considered.
However, when the Gram guidance strength is adjusted to account for increased estimation uncertainty, the feasible operating region expands substantially.
In particular, we observe that Gram-guided diffusion remains consistently superior to the likelihood-guided DM  even for very small data block lengths, with values as low as $N_d=20$ remaining effective across the considered SNR range.

This behavior demonstrates that coherence-time limitations primarily constrain the allowable guidance strength rather than imposing a hard lower bound on the number of available data symbols.
In the next set of experiments, we further illustrate this effect by directly comparing the NMSE performance of the proposed method with the likelihood-guided DM for extremely small values of $N_d$.

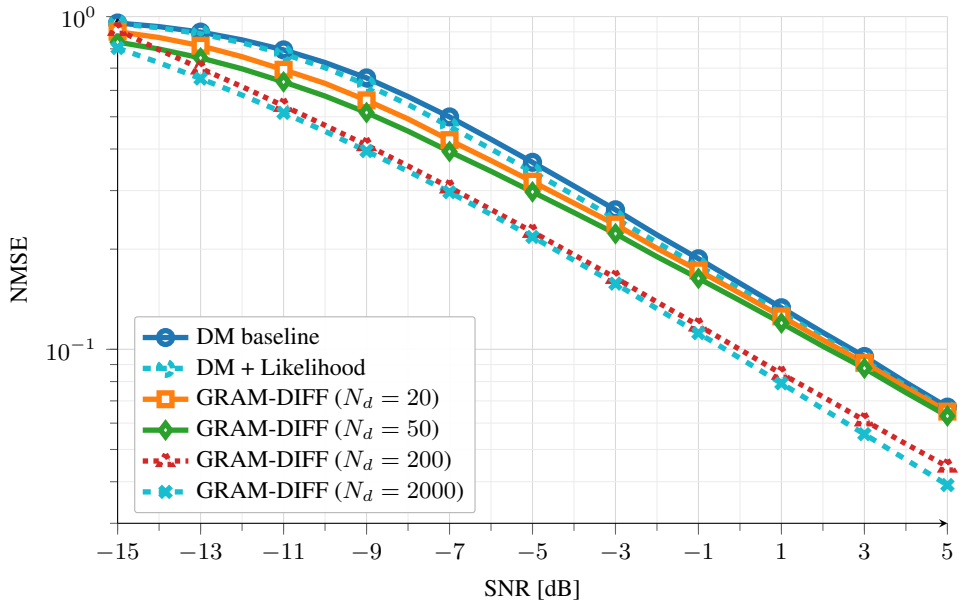
\begin{figure}[t]
  \centering
    \begin{tikzpicture}
        \begin{axis}[
          width=12.5cm,height=8.3cm,
          xmin=-15,xmax=5,
          ymin=3e-2,ymax=1,
          ymode=log,
          axis lines=left,
          xlabel={SNR [dB]},
          ylabel={NMSE},
          label style={font=\small},
          tick label style={font=\small},
          tick align=outside,
          tick style={line width=0.4pt, black!60},
          xtick={-15,-13,...,5},
          minor xtick={-15,-14,...,5},
          grid=both,
          major grid style={line width=0.2pt, draw=black!15},
          minor grid style={line width=0.15pt, draw=black!8},
          legend style={
            at={(0.02,0.02)},
            anchor=south west,
            draw=black!30,
            fill=white,
            fill opacity=0.95,
            text opacity=1,
            font=\small,
            rounded corners=1pt,
            inner sep=2pt,
          },
          legend cell align=left,
        ]

        \addplot+[color=cDM, line width=2pt, mark=o, mark size=2.6pt, mark repeat=2]
        coordinates {
        (-15,0.956674993) (-14,0.932828784) (-13,0.898368776) (-12,0.851259053) (-11,0.79441011)
        (-10,0.727668762) (-9,0.654289067) (-8,0.574952424) (-7,0.499485046) (-6,0.427704483)
        (-5,0.364863485) (-4,0.309671074) (-3,0.262683958) (-2,0.220814109) (-1,0.1872347)
        (0,0.157922253) (1,0.133347675) (2,0.112339757) (3,0.095159866) (4,0.079534099)
        (5,0.067043602)
        };
        \addlegendentry{DM baseline}
        
        \addplot+[color=cLIKE, line width=2pt, mark=o, mark size=2.6pt, mark repeat=2, dashed]
        coordinates {
        (-15,0.9525881) (-14,0.9248568) (-13,0.8880149) (-12,0.8341098) (-11,0.7732138)
        (-10,0.7018799) (-9,0.6236374) (-8,0.5442631) (-7,0.4674852) (-6,0.3998596)
        (-5,0.3414079) (-4,0.2907076) (-3,0.2473322) (-2,0.2094728) (-1,0.1783003)
        (0,0.1516210) (1,0.1287495) (2,0.1088099) (3,0.09236980) (4,0.07785374)
        (5,0.06560829)
        };
        \addlegendentry{DM + Likelihood}
        
        \addplot+[
          color=cGRAM, line width=2pt,
          mark=square*, mark size=2.4pt, mark repeat=2,
          mark options={fill=white, draw=cGRAM}
        ]
        coordinates{
        (-15,0.9006971) (-14,0.8652440) (-13,0.8172359) (-12,0.7576549) (-11,0.6923343)
        (-10,0.6299376) (-9,0.5601377) (-8,0.4925341) (-7,0.4254167) (-6,0.3691521)
        (-5,0.3190675) (-4,0.2749634) (-3,0.2371442) (-2,0.2011513) (-1,0.1725557)
        (0,0.1474491) (1,0.1255727) (2,0.1066310) (3,0.0910800) (4,0.0766909)
        (5,0.06500658)
        };
        \addlegendentry{GRAM-DIFF ($N_d=20$)}

        
        \addplot+[
          color=cORACLE, line width=2pt,
          mark=diamond*, mark size=2.5pt, mark repeat=2,
          mark options={fill=white, draw=cORACLE}
        ]
        coordinates{
        (-15,0.8380436) (-14,0.7988856) (-13,0.7514812) (-12,0.6967175) (-11,0.6365952)
        (-10,0.5770430) (-9,0.5146220) (-8,0.4526191) (-7,0.3930235) (-6,0.3429396)
        (-5,0.2972939) (-4,0.2576286) (-3,0.2226392) (-2,0.1898958) (-1,0.1633800)
        (0,0.1401420) (1,0.1199292) (2,0.1021347) (3,0.08772904) (4,0.07413346)
        (5,0.06304427)
        };
        \addlegendentry{GRAM-DIFF ($N_d=50$)}

        \addplot+[
          color=cEST, line width=2pt, dotted,
          mark=triangle*, mark size=2.6pt, mark repeat=2,
          mark options={fill=white, draw=cEST}
        ]
        coordinates{
        (-15,0.9125285) (-14,0.7947381) (-13,0.6987044) (-12,0.6158683) (-11,0.5393007)
        (-10,0.4712543) (-9,0.4109491) (-8,0.3557130) (-7,0.3066213) (-6,0.2631831)
        (-5,0.2254651) (-4,0.1923650) (-3,0.1639997) (-2,0.1387690) (-1,0.1179855)
        (0,0.09989185) (1,0.08444799) (2,0.07199838) (3,0.06117133) (4,0.05197394)
        (5,0.04443048)
        };
        \addlegendentry{GRAM-DIFF ($N_d=200$)}

        \addplot+[
          color=cLIKE, line width=2pt, dashed,
          mark=x, mark size=2.9pt, mark repeat=2
        ]
        coordinates{
        (-15,0.8046212) (-14,0.7268641) (-13,0.6512193) (-12,0.5809162) (-11,0.5137955)
        (-10,0.4520712) (-9,0.3944145) (-8,0.3428931) (-7,0.2966368) (-6,0.2543182)
        (-5,0.2176813) (-4,0.1852136) (-3,0.1575337) (-2,0.1325100) (-1,0.1118462)
        (0,0.09405626) (1,0.07904492) (2,0.06623647) (3,0.05558591) (4,0.04664312)
        (5,0.03911159)
        };
        \addlegendentry{GRAM-DIFF ($N_d=2000$)}

        \end{axis}
        \end{tikzpicture}
  \caption{3GPP: NMSE versus SNR under coherence-time limited Gram estimation.}
  \label{fig:robust_nd_small}
\end{figure}

Fig.~\ref{fig:robust_nd_small} compares GRAM-DIFF with the diffusion prior baseline and the DM+likelihood reference under coherence-time limited data block lengths.
Even with only $N_d=20$ data symbols, Gram-guided diffusion consistently outperforms the baseline across the entire SNR range from $-15$ to $5$~dB, confirming that the proposed framework remains effective under severe sample limitations.

To quantify the tolerance of the reverse diffusion dynamics to Gram estimation errors, we further include an intermediate regime with $N_d=200$ and a large-sample reference using $N_d=2000$.
The relatively small gap between the $N_d=200$ and $N_d=2000$ curves indicates that the proposed method is robust to moderate inaccuracies in the estimated Gram matrix $\widehat{\mathbf R}$, and does not rely on highly precise second-order statistics to achieve most of its performance gain.
Instead, Gram guidance primarily acts as a controllable structural bias that stabilizes the reverse diffusion trajectory, while the remaining benefit of further improving $\widehat{\mathbf R}$ diminishes once the dynamics enter the stable operating regime.

\vspace{0.5em}
\noindent
\textbf{Continuous Performance Interpolation and Vanishing-Guidance Limit.}
An important empirical observation is that Gram-guided diffusion does not exhibit a hard performance failure threshold with respect to $N_d$ when the guidance strength is properly scaled.
As the data block length decreases and the Gram estimation error increases, reducing the guidance coefficient $\lambda_{\mathrm{G}}$ effectively suppresses the impact of unreliable structural information.
In the extreme sample-limited regime (e.g., $N_d=5$), we observe that suitably scaled Gram guidance can still provide consistent performance improvements over the DM+likelihood model, whereas overly aggressive guidance may introduce biased update directions and degrade performance.
Moreover, when $\lambda_{\mathrm{Gram}}$ is gradually annealed together with decreasing $N_d$, the resulting performance curve smoothly converges to that of the diffusion model with likelihood guidance only.
Since the likelihood-guided DM model itself consistently outperforms the diffusion prior baseline, this property guarantees that Gram-guided diffusion does not collapse below the baseline performance even in extremely sample-limited regimes.

\section{Appendix}
\label{app:impl}

\subsection{Implementation Details: Stability of Gram Guidance}
\label{sec:cov_stability}

When applying gram matrix guidance, the resulting update may exhibit a large magnitude due to the global, second-order nature of the gram matrix loss.
To ensure numerical stability of the reverse diffusion process, we restrict the per-step gram guidance update using a norm-based clipping strategy.
Specifically, let $\Delta \mathbf{x}_{\mathrm{gram},t}$ denote the gram matrix guidance update at diffusion step $t$.
We apply the following clipping operation on a per-sample basis:
\begin{equation}
\Delta \mathbf{x}_{\mathrm{gram},t} \leftarrow 
\Delta \mathbf{x}_{\mathrm{gram},t}
\cdot
\min\!\left(
1,
\frac{Th}{\|\Delta \mathbf{x}_{\mathrm{gram},t}\|_2 + \varepsilon}
\right),
\end{equation}
where $Th$ is a predefined clipping threshold and $\varepsilon$ is a small constant for numerical stability.

This norm-based clipping preserves the direction of the gram matrix guidance update while limiting its overall magnitude.
In practice, we found this strategy to be sufficient for stabilizing the reverse diffusion trajectory across all tested SNR regimes.

\subsection{Additional Results: Coherence-Time Constraints}
\label{app:extreme_nd}

\begin{figure}[ht]
  \centering
    \begin{tikzpicture}
        \begin{axis}[
          width=12.5cm,height=8.3cm,
          xmin=-15,xmax=5,
          ymin=3e-2,ymax=1,
          ymode=log,
          axis lines=left,
          xlabel={SNR [dB]},
          ylabel={NMSE},
          label style={font=\small},
          tick label style={font=\small},
          tick align=outside,
          tick style={line width=0.4pt, black!60},
          xtick={-15,-13,...,5},
          grid=both,
          major grid style={line width=0.25pt, draw=black!12},
          minor grid style={line width=0.15pt, draw=black!6},
          ytick={1,0.1,0.01},
          yticklabels={$10^0$,$10^{-1}$,$10^{-2}$},
          legend style={
            at={(0.98,0.98)},
            anchor=north east,
            draw=black!15,
            fill=white,
            fill opacity=0.85,
            text opacity=1,
            font=\small,
            rounded corners=2pt,
            inner sep=2pt,
          },
          legend cell align=left,
        ]

        \addplot+[
          color=cDM,
          line width=2.4pt,
          mark=o,
          mark size=2.4pt,
          mark repeat=3
        ]
        coordinates {
        (-15,0.956674993) (-14,0.932828784) (-13,0.898368776) (-12,0.851259053) (-11,0.79441011)
        (-10,0.727668762) (-9,0.654289067) (-8,0.574952424) (-7,0.499485046) (-6,0.427704483)
        (-5,0.364863485) (-4,0.309671074) (-3,0.262683958) (-2,0.220814109) (-1,0.1872347)
        (0,0.157922253) (1,0.133347675) (2,0.112339757) (3,0.095159866) (4,0.079534099)
        (5,0.067043602)
        };
        \addlegendentry{DM baseline}
        
        \addplot+[
          color=cLIKE,
          line width=2.3pt,
          dashed,
          opacity=0.85,
          mark=*,
          mark size=1.6pt,
          mark repeat=4
        ]
        coordinates {
        (-15,0.9525881) (-14,0.9248568) (-13,0.8880149) (-12,0.8341098) (-11,0.7732138)
        (-10,0.7018799) (-9,0.6236374) (-8,0.5442631) (-7,0.4674852) (-6,0.3998596)
        (-5,0.3414079) (-4,0.2907076) (-3,0.2473322) (-2,0.2094728) (-1,0.1783003)
        (0,0.1516210) (1,0.1287495) (2,0.1088099) (3,0.09236980) (4,0.07785374)
        (5,0.06560829)
        };
        \addlegendentry{DM + Likelihood}
        
        \addplot+[
          color=cGRAM,
          line width=2.2pt,
          solid,
          mark=square*,
          mark size=2.2pt,
          mark repeat=3,
          mark options={fill=white}
        ]
        coordinates{
        (-15,0.9006971) (-14,0.8652440) (-13,0.8172359) (-12,0.7576549) (-11,0.6923343)
        (-10,0.6299376) (-9,0.5601377) (-8,0.4925341) (-7,0.4254167) (-6,0.3691521)
        (-5,0.3190675) (-4,0.2749634) (-3,0.2371442) (-2,0.2011513) (-1,0.1725557)
        (0,0.1474491) (1,0.1255727) (2,0.1066310) (3,0.0910800) (4,0.0766909)
        (5,0.06500658)
        };
        \addlegendentry{GRAM-DIFF ($N_d=20$)}

        \addplot+[
          color=cGRAM,
          line width=2.0pt,
          densely dashed,
          opacity=0.85,
          mark=triangle*,
          mark size=2.1pt,
          mark repeat=4,
          mark options={fill=white}
        ]
        coordinates{
        (-15,0.9315830) (-14,0.8975795) (-13,0.8519053) (-12,0.7940069) (-11,0.7280564)
        (-10,0.6618107) (-9,0.5889561) (-8,0.5171627) (-7,0.4446036) (-6,0.3837621)
        (-5,0.3288639) (-4,0.2823701) (-3,0.2426762) (-2,0.2047877) (-1,0.1750540)
        (0,0.1492365) (1,0.1270608) (2,0.1076127) (3,0.09177936) (4,0.07711521)
        (5,0.06522168)
        };
        \addlegendentry{GRAM-DIFF ($N_d=10$)}

        \addplot+[
          color=cGRAM,
          line width=1.8pt,
          dotted,
          opacity=0.65,
          mark=diamond*,
          mark size=2.0pt,
          mark repeat=5,
          mark options={fill=white}
        ]
        coordinates{
        (-15,0.9396965) (-14,0.9097712) (-13,0.8663787) (-12,0.8114359) (-11,0.7460909)
        (-10,0.6780590) (-9,0.6031329) (-8,0.5282335) (-7,0.4534096) (-6,0.3902963)
        (-5,0.3337905) (-4,0.2856014) (-3,0.2449026) (-2,0.2069611) (-1,0.1766663)
        (0,0.1501513) (1,0.1277199) (2,0.1082567) (3,0.09227099) (4,0.07747461)
        (5,0.06541284)
        };
        \addlegendentry{GRAM-DIFF ($N_d=5$)}

        \end{axis}
        \end{tikzpicture}
  \caption{NMSE versus SNR under extreme coherence-time constraints (3GPP channel). We compare the diffusion prior baseline (DM), the DM+likelihood reference, and GRAM-DIFF with very small data block lengths ($N_d=5,10,20$).}
  \label{fig:robust_nd_extreme}
\end{figure}
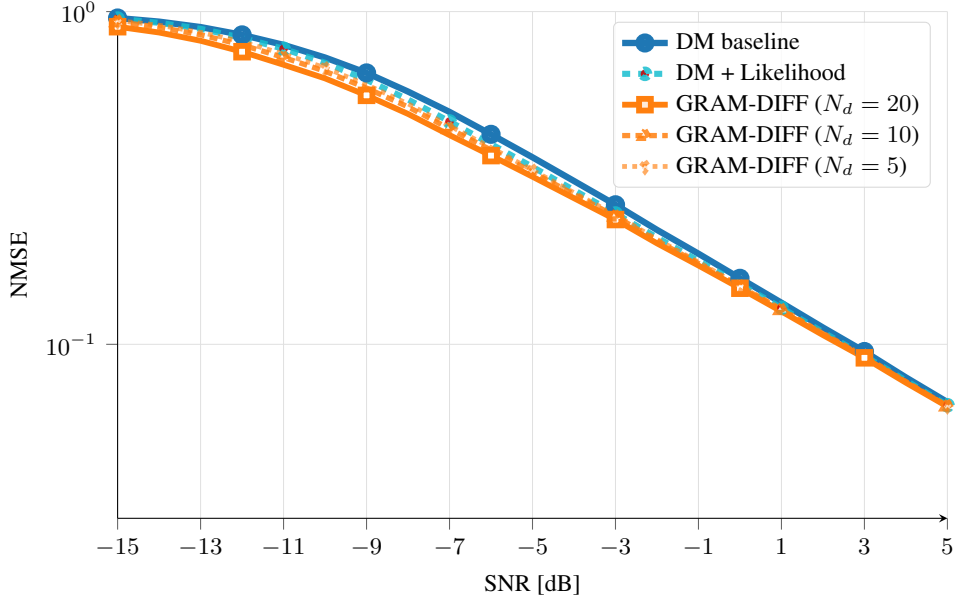

To complement Fig.~\ref{fig:robust_nd_small}, we report additional ablation results under extremely short coherence-time regimes with $N_d\in\{5,10,20\}$ in Fig.~\ref{fig:robust_nd_extreme}.
These regimes correspond to highly noisy Gram estimation, where Gram guidance may become unstable.
In contrast, by adapting the Gram guidance strength $\lambda_{\mathrm{Gram}}$ according to the reliability of $\widehat{\mathbf R}$, GRAM-DIFF remains operationally effective even when only a handful of data symbols are available.

As shown in Fig.~\ref{fig:robust_nd_extreme}, the performance curves for $N_d=5$ and $N_d=10$ remain close to the $N_d=20$ case, especially in moderate-to-high SNR regimes, while still maintaining a consistent gain over the DM+likelihood reference.
Moreover, as $N_d$ decreases, the benefit contributed by Gram guidance gradually diminishes and the overall performance smoothly approaches that of the likelihood-guided diffusion sampler, which corroborates the vanishing-guidance interpretation discussed in Sec.~\ref{sec:robust}.

\newpage

\bibliography{ref}
\bibliographystyle{ieeetr}

\end{document}